\documentclass{emulateapj}

\shorttitle{Star formation in the extreme outer Galaxy}
\shortauthors{Izumi et al.}

\begin{document}
\title{Discovery of star formation in the extreme outer Galaxy possibly induced by a high-velocity cloud impact}

\author{NATSUKO IZUMI\altaffilmark{1,2}, NAOTO KOBAYASHI\altaffilmark{1}, CHIKAKO YASUI\altaffilmark{2},
 ALAN T. TOKUNAGA\altaffilmark{3}, MASAO SAITO\altaffilmark{4,5}, AND SATOSHI HAMANO\altaffilmark{1,2}}

\altaffiltext{1}{Institute of Astronomy, School of Science, The University of Tokyo, 2-21-1 Osawa, Mitaka, Tokyo 181-0015,
Japan; izumi@ioa.s.u-tokyo.ac.jp}
\altaffiltext{2}{Department of Astronomy, Graduate School of Science, The University of Tokyo, 7-3-1 Hongo, Bunkyo-ku, Tokyo 113-0033, Japan}
\altaffiltext{3}{Institute for Astronomy, University of Hawaii, 2680 Woodlawn Drive, Honolulu, HI 96822, USA}
\altaffiltext{4}{National Astronomical Observatory of Japan, 2-21-1 Osawa, Mitaka, Tokyo 181-8588, Japan}
\altaffiltext{5}{Joint ALMA Observatory, Ave. Alonso de Cordova 3107, Vitacura, Santiago, Chile}

\begin{abstract}
We report the discovery of star formation activity in perhaps the most distant molecular cloud in the extreme outer galaxy.
 We performed deep near-infrared imaging with the Subaru 8.2 m telescope, and found two young embedded clusters at two CO peaks of
  ``Digel Cloud 1" at the kinematic distance of $D$ = 16 kpc (Galactocentric radius $R_{\rm G}$ = 22 kpc).
 We identified 18 and 45 cluster members in the two peaks, and the estimated stellar densities are $\sim 5$ and $\sim 3$ pc$^{-2}$, respectively. 
 The observed $K$-band luminosity function suggests that the age of the clusters is less than 1 Myr and also that the distance to the clusters
 is consistent with the kinematic distance.
 On the sky, Cloud 1 is located very close to the \ion{H}{1} peak of high-velocity cloud Complex H,
 and there are some \ion{H}{1} intermediate velocity structures between the Complex H and the Galactic disk, which could indicate an interaction between them.
 We suggest the possibility that Complex H impacting on the Galactic disk has triggered star formation in Cloud 1 as well as the formation of the Cloud 1 molecular cloud.
\end{abstract}

\section{Introduction}
The extreme outer galaxy (EOG), which we define as the region with a Galactocentric radius ($R_{\rm G}$) of more than
 18 kpc \citep{Kobayashi08,Yasui08}, has a very different environment from the solar neighborhood with a much lower gas density \citep{Wolfire03},
 lower metallicity \citep{Smartt97}, and little or no perturbation from the spiral arms.
 Such a region not only defines the size and shape of our Galaxy,
 but it also serves as an excellent laboratory for studying the star-forming process in low-density and low-metallicity environments.
 Because such environments may have similar characteristics that existed in the early phase of Galaxy formation \citep{Kobayashi08},
 we might be able to directly observe the galaxy formation processes in unprecedented detail at a much closer distance than distant galaxies.

The Digel Clouds were discovered by the very first survey of molecular clouds in the EOG \citep{Digel94} and are some of the best targets for studying star formation in the EOG.
 The Digel Clouds are composed of eight molecular clouds (Cloud 1--8), and $R_{\rm G}$ is estimated at more than 20 kpc for many of them from the kinematic distance.
 A star-forming region has been identified in Cloud 2 \citep{De93,Kobayashi00,Yasui06,Yasui08}, which has the highest CO luminosity among all Digel Clouds.
These clouds are a very valuable sample, because the expected number of molecular clouds in such a low-density region is very small \citep{Snell02}.
 
In this paper, we report the discovery of star-forming clusters in Cloud 1, which has the largest dynamical mass among all of the Digel Clouds.
 Despite a relatively strong $^{13}$CO line, no star-forming activity has been reported thus far because of the very large distance with perhaps
 the largest Galactocentric radius ($R_{\rm G}$ $\approx$ 22 kpc)
 among all known clouds in the Galaxy\footnote{Although the kinematic distance of Cloud 2 is larger than that of Cloud 1 in 
 Digel et al.'s (1994) original list, the photometric distance of Cloud 2 is found to be smaller than the kinematic distance ($R_{\rm G} \approx$ 19 kpc). See Section 4 for more detail.}.
 In view of the close coincidence on the sky with the high-velocity cloud (HVC) Complex H,
 we suggest that star and molecular cloud formation in Cloud 1 is triggered by HVC accretion onto the Galactic disk.

\begin{figure*}
\epsscale{1.0}
\plotone{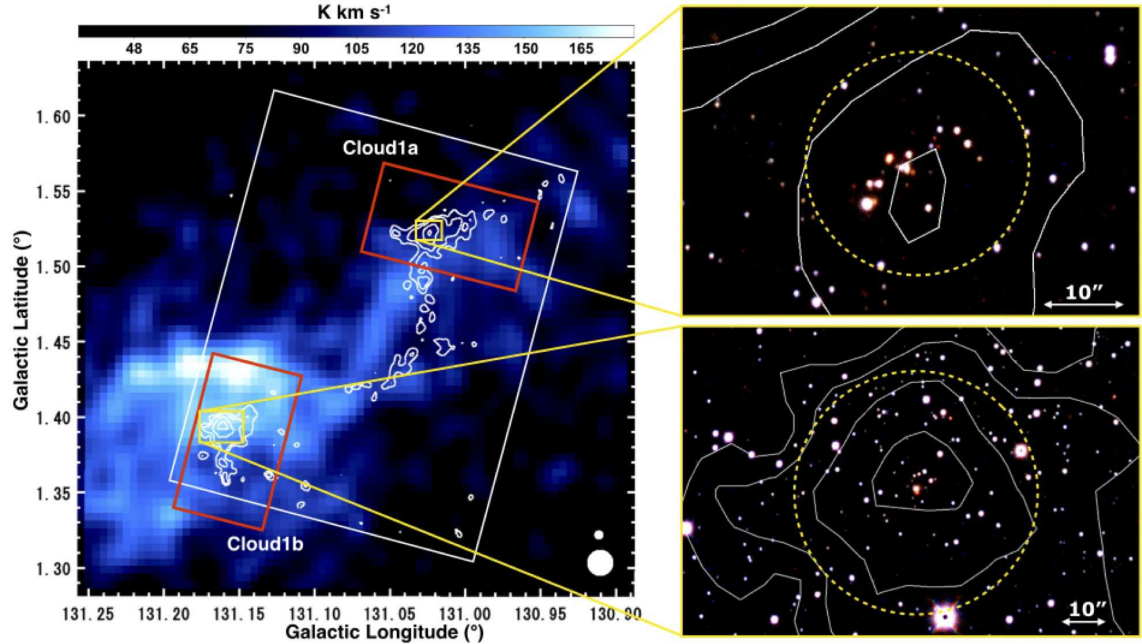}
\caption{Left: \ion{H}{1} map around Cloud 1 (from CGPS data : $v_{\rm LSR} = -104.5\sim -99.6$ km s$^{-1}$ ).
The white contours show the integrated $^{12}$CO (1--0) map of Cloud 1 (from our NRO 45 m telescope data: $v_{\rm LSR} = -104.1\sim -99.1$ km s$^{-1}$), and contour levels are at 2$\sigma$, 3$\sigma$, 5$\sigma$, 7$\sigma$ (1$\sigma$ = 0.5 K km s $^{-1}$).
The white and red boxes show the mapping size of the NRO 45 m observation and the field of view of the Subaru MOIRCS ($4^\prime \times 7^\prime$), respectively.
The large and small white filled circles at the lower right corner show the beam sizes of the DRAO ($\sim$58$^{\prime \prime}$) and NRO 45 m telescope ($\sim$17$^{\prime \prime}$), respectively.
Right: $JHK$ pseudo color images of the Cloud 1 clusters from our Subaru MOIRCS data.
The white contours show the same as the $^{12}$CO map as in the left panel.
The yellow dotted circles show the defined cluster regions (radius of Cloud 1a circle : 14$^{\prime \prime}$, Cloud 1b : 28$^{\prime \prime}$).} 
 \label{fig1}
\end{figure*}
\begin{table*}
\begin{center}
\caption{Parameters of the Observations Containing the Region Cloud 1 or Complex H}
\begin{tabular}{ccccccc}
\hline  \hline 
Project & Telescope & Wavelength & Coverage & Velocity Range& Velocity Resolution& Resolution\tablenotemark{a} \\
&&&&(km s$^{-1}$)&(km s$^{-1}$)& \\ \tableline \tableline
CGPS             &DRAO&21.1 cm (\ion{H}{1})&$74^\circ.2<l<147^\circ.3$ &-150 to 50&1.32&$58^{\prime \prime} \times 58^{\prime \prime} {\rm cosec }\delta$\\
&Synthesis Telescope&                    &$-3^\circ.6<b<+5^\circ.6 $ &                  &        &\\ \hline
LAB survey&Villa Elisa&21.1 cm (\ion{H}{1})&$0^\circ<l<360^\circ$&-450 to 450&1.27&$30^{\prime}.0$\\
                   & 30 m       &                           &$-90^\circ<b<-25^\circ$&               &       &                          \\
                   &Dwingeloo&21.1 cm (\ion{H}{1})&$0^\circ<l<360^\circ$&-450 to 450&1.25&$35^{\prime}.7$\\
                   & 25 m       &                           &$-30^\circ<b<+90^\circ$&               &       &                          \\     \hline
FCRAO outer&FCRAO&2.6 mm ($^{12}$CO)&$102^\circ.49<l<141^\circ.54$ &-153 to 40&0.98&$45^{\prime \prime}$\\
Galaxy Survey                       & 14 m               &           &$-3^\circ.03<b<+5^\circ.41 $ &                  &        &\\ \hline                   
$WISE$          &$WISE$     &3.4 $\mu$m      & All-sky                    & $\cdots$                    &  $\cdots$     &$6^{\prime \prime}.1$\\
                   & satellite               &4.6 $\mu$m      & All-sky                    & $\cdots$                    &  $\cdots$     &$6^{\prime \prime}.4$\\ 
                   &                &12  $\mu$m      & All-sky                    & $\cdots$                    &  $\cdots$     &$6^{\prime \prime}.5$\\
                   &                &22  $\mu$m      & All-sky                    & $\cdots$                    &  $\cdots$     &$12^{\prime \prime}.0$\\  \hline
IPHAS        &Issac Newton 2.5 m&656.8 nm (H$\alpha$)     &$29^\circ<l<215^\circ$& $\cdots$                    &  $\cdots$     & 0$^{\prime \prime}.333$  ${\rm pixel}^{-1}$\\
                   &(Wide Field Camera)            &                                       &$-5^\circ<b<+5^\circ$  &                     &       &                                                           \\ \hline
This paper&Subaru 8.2 m&1.25  $\mu$m ($J$) &$4^\prime \times 7^\prime \times 2$& $\cdots$ &  $\cdots$ & 0$^{\prime \prime}.112 $  ${\rm pixel}^{-1}$\\
                    & (MOIRCS)   &1.65  $\mu$m ($H$) &$4^\prime \times 7^\prime \times 2$& $\cdots$ &  $\cdots$ & 0$^{\prime \prime}.112 $  ${\rm pixel}^{-1}$\\
                    &                     &2.15  $\mu$m ($K_{\rm S}$) &$4^\prime \times 7^\prime \times 2$& $\cdots$ &  $\cdots$ & 0$^{\prime \prime}.112 $  ${\rm pixel}^{-1}$\\            
                    &NRO45 m&2.6 mm ($^{12}$CO)&$15^\prime \times 16^\prime$&-110 to -90&0.25&$\sim$ 17$^{\prime \prime}$\\                                                                           
\tableline \tableline
\end{tabular}
\label{tbl:table1}
\tablecomments{$^{a}$ The meaning of the resolution with the IPHAS and Subaru 8.2 m observation is pixel size, and others is diffraction limited resolution.}
\end{center}
\end{table*}
\begin{figure}
\epsscale{1.0}
\plotone{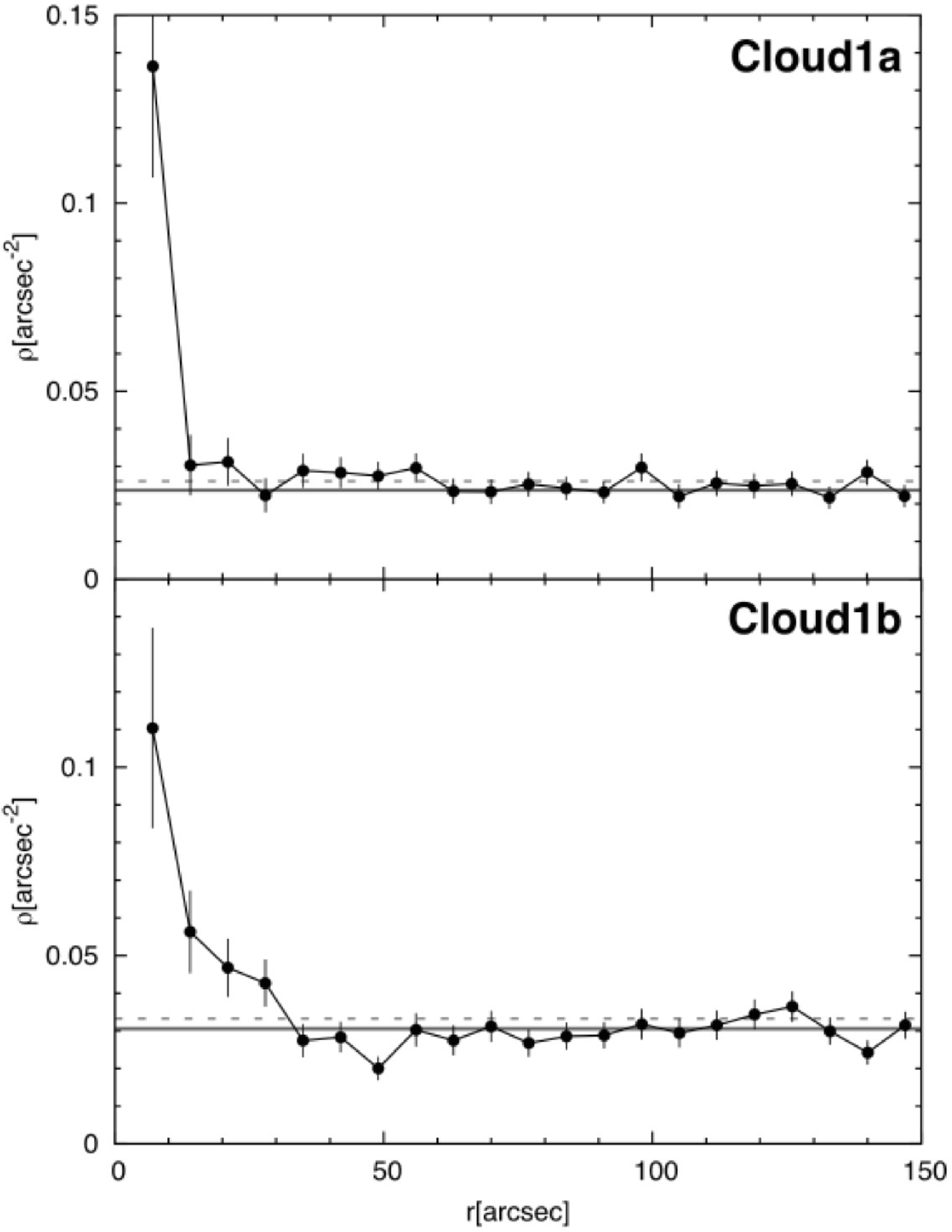}
\caption{Radial variation of the projected stellar number density of stars around the center of the CO peaks.
The center position of Clouds 1a and 1b are ($l,b$)=($131^\circ.02$, $1^\circ.52$) and ($131^\circ.16$, $1^\circ.39$), respectively.
The error bars represent Poisson errors (1$\sigma$).
The horizontal solid lines and horizontal dashed lines indicate the average density of stars outside the Cloud 1 clouds and the Poisson errors (3$\sigma$), respectively.}
\label{fig2}
\end{figure}
\section{Observations and data reduction}
\subsection{Subaru MOIRCS Imaging}
We obtained $J$ (1.25 $\mu$m)-, $H$ (1.65 $\mu$m)-, and $K_{\rm S}$ (2.15 $\mu$m)-band deep images of the two CO peaks of Cloud 1 (Cloud 1a and  Cloud 1b).
 The observations were conducted on 2006 September 2 UT with a wide-field near-infrared (NIR) camera, MOIRCS \citep{Ichikawa06} on the Subaru 8.2 m telescope.
 It provides a $4^\prime \times 7^\prime$ field of view with a $0^{\prime \prime}.117$ pixel$^{-1}$ scale and employs the Mauna Kea Observatory (MKO) NIR photometric filters \citep{Tokunaga02}.
 The total integration time was $\sim$700, 600, and 600 s for $J$, $H$, and $K_{\rm S}$ bands, respectively.
 The observing condition was photometric and the seeing was excellent ($\sim 0^{\prime \prime}.4$) throughout the observing period.
 
\subsection{Data Reduction}
All of the data for each band were reduced with IRAF ver 2.14 with standard procedures: dark subtraction, flat fielding,
 bad-pixel correction, median-sky subtraction, image shifts with dithering offsets, and combining.
 The stellar FWHM in final images of the $J$, $H$, and $K_{\rm S}$ bands are $0^{\prime \prime}.45$, $0^{\prime \prime}.43$, and $0^{\prime \prime}.40$, respectively. 
 $JHK_{\rm S}$ photometry has been performed using the IRAF APPHOT package, with aperture diameters of $1^{\prime \prime}.17$ (10 pixels).
 The aperture sizes were chosen to achieve high signal-to-noise ratio (S/N) and sufficient flux count from the stellar objects. 
 Photometric calibration was done using the standard star Persson9166 \citep[GSPC P330-E; $J$ = 11.772, $H$ = 11.455, $K$ = 11.419;][]{Leggett06}.
 The resultant 10$\sigma$ limiting magnitudes in the $J$, $H$, and $K_{\rm S}$ bands were estimated at 21.0, 20.5, and 19.5 mag, respectively.
 
\subsection{CO Data with the Nobeyama Radio Observatory (NRO) 45 m Telescope}
We have performed observations of Cloud 1 in the CO ($J$= 1--0; 115.271 GHz) line with the NRO 45 m telescope\footnote{The Nobeyama Radio 
Observatory is a branch of the National Astronomical Observatory of Japan, National Institutes of Natural Sciences.} in 2007 December.
 We used the 25-BEam Array Recceiver System (BEARS) in the double-side band mode, which has 5 $\times$ 5 beams separated by 41$^{\prime \prime}$.1 on the plane of the sky
 \citep{Sunada00,Yamaguchi00}. The telescope beamwidth and the main beam efficiency at 115 GHz was 14$^{\prime \prime}$.5 and 0.39, respectively.
 As a backend, the autocorrelator was adopted, and the typical noise level was 0.40 K in $T^{*}_{\rm A}$ with 0.25 km s$^{-1}$ resolution in CO.
   
We employed an on-the-fly mode developed for the NRO 45 m telescope \citep{Sawada08}.
 The data sampling interval of the R.A. or decl. scans is $\sim$ 1$^{\prime \prime}$, and the separation between the each scans is 5$^{\prime \prime}$.1.
 After subtracting linear baselines, the data were convolved with a Gaussian-tapered Bessel function whose FWHM was 14$^{\prime \prime}$ and
 resampled onto a 6$^{\prime \prime}$ grid. Since the telescope beam is a Gaussian with an FWHM of 14$^{\prime \prime}$.5--14$^{\prime \prime}$.8, and effective FWHM resolution of $\sim$17$^{\prime \prime}$.
 Finally, to reduce the ``scanning effect'', where some of the conditions outlined above have not been adequately satisfied, resulting in an effective noise level on the final map 
 higher than the theoretical value, we combine the two maps scanned R.A. and decl. directions using the basket-weaving method \citep{Emerson88}. 
 We mapped an area of 15$^{\prime}$ $\times$ 16$^{\prime}$ in CO to cover the entire Cloud 1 (Figure \ref{fig1}).
 
 The atmospheric corrected temperature scale $T^{*}_{\rm A}$ is obtained with the chopper wheel method. 
 During the observation, the typical system noise temperature of BEARS in the double-side band was 400 K at CO frequencies. 
 The telescope pointing was checked about every 90 minutes by five-point scans of the SiO maser source S-Per [R.A. = 02$^{\rm h}$ 22$^{\rm m}$ 51$^{\rm s}$.713, decl. = 58$^{\circ}$ 35$^{\prime}$ 11$^{\prime \prime}$. 50 (J2000)]
 with SIS 49 GHz receiver (S40). The measured pointing errors ranged from 1$^{\prime \prime}$.5 to 6$^{\prime \prime}$.0 during the observing run.
 
\subsection{Previous Observations}
In addition to the our NIR and CO observation, we used archived data from previous observations ``the Canadian Galactic Plane Survey (CGPS)"\citep{Tayer03},
 ``the FCRAO outer Galaxy Survey" \citep{Heyer98}, ``the Leiden/Argentine/Bonn (LAB) Survey of Galactic \ion{H}{1}''\citep{Kalberla05},
 ``the $Wide$-$field$ $Infrared$ $Survey$ $Explorer$ ($WISE$) all sky survey" \citep{Wright10}, and ``the INT/WFC Photometric H$\alpha$ Survey (IPHAS)'' \citep{Drew05}.
 We summarize in Table 1 the observations we used to study Cloud 1 or Complex H in this paper. 
 
\section{Results}

\subsection{Detection of Star-forming Regions in Cloud 1} 
In the false-color pictures (Figure \ref{fig1}), we found two star clusters of reddened stars, in the vicinity of the two CO peaks, Cloud 1a and 1b \citep{Digel94}.
 We examined the positional relationship between these clusters, Cloud 1, and the other foreground molecular clouds on the sky to confirm that Cloud 1 is the only molecular cloud positionally coincident with the clusters.
 Therefore, we concluded that these clusters are embedded clusters associated with Cloud 1.
 In view of the location of the clusters in the molecular clouds, near the peak of the dense CO core, these clusters are likely to be younger than $\sim$ 3 Myr \citep{Lada03}.
 
In order to define the extent of the clusters, we derived the stellar number density distribution around the CO peaks (Figure \ref{fig2}).
 As a result, we defined the radius of the Cloud 1a and 1b cluster region to be $14^{\prime \prime}$ and $28^{\prime \prime}$, respectively.
 
\subsection{Identification of Cluster Members}
\begin{figure}
\epsscale{1.0}
\plotone{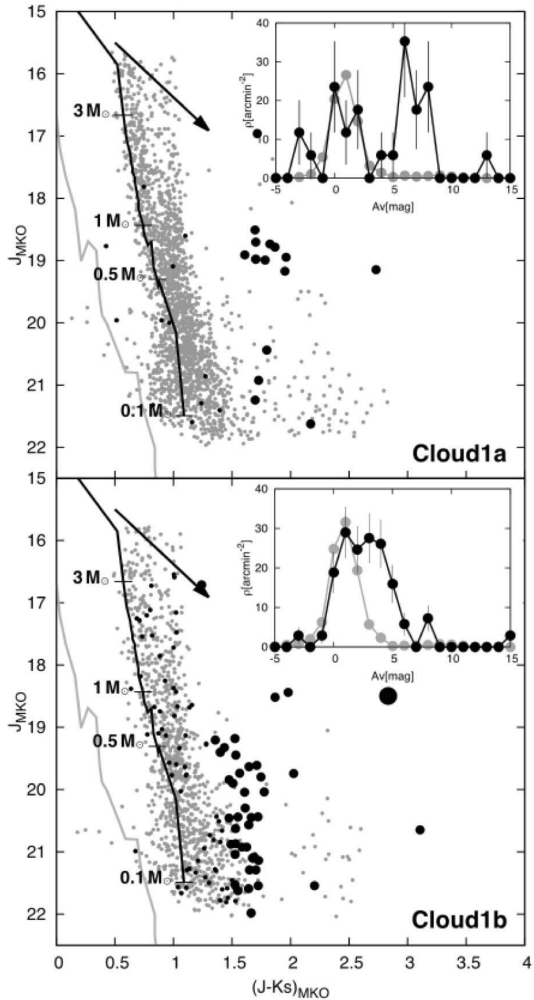}
\caption{$(J-K_{\rm S})$ vs. $J$ color-magnitude diagram for the Cloud 1 clusters in the MKO system.
 The top-right inset of each panel shows $A_V$ distributions of stars in the cluster regions (black filled circles and lines) and in the field (gray filled circles and lines).
 The error bars show the uncertainties assuming Poisson statistics.
 The gray lines show dwarf tracks \citep{Bessel88}, while the black lines show isochrone models (1.0 Myr; Siess et al. 2000).
 The black arrows show the redding vectors of $A_{V}$ = 5 mag.
 The filled circles show stars in the cluster regions of Cloud 1a and Cloud 1b, respectively,
 small circles show field stars in the cluster regions, large circles show identified cluster members,
 and the very large circle shows the most luminous dereddened source.
 The gray dots show field stars in the field of view.}
\label{fig3}
\end{figure}

We identified cluster members using the $J - K_{\rm S}$ versus $J$ color-magnitude (CM) diagram of all the detected sources with S/N $> 5\sigma$ in all three bands (Figure \ref{fig3})
 following the method used in \citet{Yasui06,Yasui08}.
 On the CM diagram, we estimated the nominal extinction ($A_V$) of all the detected sources by measuring the distance along the reddening vector \citep{Rieke85}
 from a reference isochrone of an assumed age of 1 Myr and a kinematic distance of $D$ = 16 kpc (see Figure \ref{fig3}).
 We found that stars with large extinction, $A_V\ge$ 4 mag and 3 mag are concentrated on the Cloud 1a and Cloud 1b cluster area, respectively,
 while stars with small extinction ($A_V =$ 0--2 mag) are uniformly distributed over the observed field (see insets in Figure \ref{fig3}).
 Therefore, we identified the cluster members with the following criteria:
 (1) distributed in the regions of Cloud 1a and Cloud 1b cluster region (see Figure \ref{fig1}) and (2) $A_V\ge$ 4 and 3 mag for Cloud 1a and Cloud 1b, respectively. 
 With this 2nd criterion, the contamination of field stars is estimated to be only 2 $\%$ (1a) and 10 $\%$ (1b), which is almost negligible.
 The number of resultant cluster members of Cloud 1a and Cloud 1b are 18 and 45, respectively.
 The radii of the defined cluster region in Cloud 1a and Cloud 1b are 1.1 pc (14$^{\prime \prime}$) and 2.2 pc (28$^{\prime \prime}$), respectively (see Figure \ref{fig1}).
 Therefore, the estimated stellar densities of the clusters are ~5 pc$^{-2}$ and ~3 pc$^{-2}$, respectively.
 The achieved limiting magnitudes correspond to the mass-detection limit of $<$ 1 $M_\odot$ for the kinematic distance ($D$ = 16 kpc).
 
 Figure \ref{fig4} shows the $K$-band luminosity functions (KLFs), which are the number of stars as a function of $K$-band magnitude, for the member of two Cloud 1 clusters.
 The estimated completeness limit of Cloud 1 data is about $K_{\rm S}$ = 20 mag.
 We estimated the detection completeness by the number of field stars, which rapidly decrease at $K_{\rm S} >$ 20 mag.
 The KLF of the Cloud 1a cluster shows a rather stochastic curve probably because of the small number of detected members due to the large differential extinction, or the truly small number of members.
 Therefore, we use the KLF of the Cloud 1b cluster for the present study assuming that the Cloud 1a and 1b clusters have similar properties. \\
 
 \subsection{Molecular Clouds Distribution in Cloud 1}
In the CO distribution of Cloud 1 (Figure \ref{fig5}), we detected two CO peaks (Cloud 1a and Cloud 1b) at $v_{\rm LSR} \sim$ 100 km s$^{-1}$ and a newly found bridge structure, which connects the two peaks.
 To estimate the masses of these clouds from the CO intensity $I_{\rm CO}$, we used the same mass-calibration ratio $N({\rm H_2})/I_{\rm CO}$ as \citet{Digel94} (2.3 $\times$ 10$^{20}$ cm$^{-2}$ (K km s$^{-1}$)$^{-1}$ ; the Galactic average value).
The estimated masses of Cloud 1a, Cloud 1b, and the bridge are 3.0 $\times$10$^3$, 3.5 $\times$10$^3$, and 4.5 $\times$10$^3$ $M_\odot$, respectively.
The estimated velocity width of Cloud 1a, Cloud 1b, and the bridge are 2.2, 2.4, and 1.9 km s$^{-1}$, respectively.
 The estimated radii of Cloud 1a and Cloud 1b are 5.6 and 4.2 pc, respectively, and the length of the bridge is 42 pc.
 The parameters for Cloud 1a and Cloud 1b are consistent with the results of \citet{Digel94}.

Figure \ref{fig5} also shows mid-infrared (MIR) pseudo color images from the $WISE$ data around Cloud 1.
 We confirmed that the Cloud 1 clusters are also detected in the MIR images as groups of compact reddened stellar objects.
 We found some other compact reddened stellar objects around the two CO peaks and in the bridge, which also appear to be associated with Cloud 1.
 The large diffuse reddened (12 $\mu$m) structures are considered to be the Galactic cirrus \citep{Meisner14} in the foreground.
 
 \subsection{Extinction Inside Cloud 1}
 
 We compared the extinction of the Cloud 1 clusters from our NIR data to that from the $^{13}$CO column density in the literature.
 \citet{Ruffle06} estimated the $^{13}$CO column density of Cloud 1a to be N($^{13}$CO) = $2.09 \pm 0.32 \times 10^{15}$ cm$^{-2}$, but no estimate was made for Cloud 1b. 
 From this column density, we estimated the extinction of Cloud 1a to be  $A_V =$ 4.4 mag, assuming the same dust-to-gas ratio and $^{13}$CO fractional abundance as the solar neighborhood \citep[e.g.,][]{Frerking82}.
 The total extinction of Cloud 1a from our data is $A_V = 5 \sim 8$ mag (see the top right inset of Figure \ref{fig3}), and the extinction by foreground interstellar clouds at $R_{\rm G} < 20$ kpc is estimated to be $A_V = 3 \sim 4$ mag \citep[e.g.,][]{Amores05}.
 Therefore, the extinction of the Cloud 1a clusters contributed from only the molecular cloud is estimated to be $A_V = 1 \sim 5$ mag.
 In view of the systematic uncertainties of related parameters, this value is roughly consistent with the extinction derived from the $^{13}$CO data.
 Although the dust-to-gas ratio in such low-metallicity environment compared to the solar neighborhood is of great interest, it is hard to discuss the possible difference without other independent data.
 
\subsection{\ion{H}{1} Distribution around Cloud 1}
The large-scale \ion{H}{1} distribution around Cloud 1 (Figure \ref{fig6}) shows that there is an \ion{H}{1} peak of the HVC Complex H (HVC 131+1-200) close to Cloud 1 on the sky with a
 separation of only $\sim 0.5^{\circ}$, though they are about 100 km s$^{-1}$ apart from each other in the line-of-sight velocities.
 Furthermore, there is a large \ion{H}{1} shell at $v_{\rm LSR}\sim -100$ km s$^{-1}$, which was originally identified by \citet{Heiles79}.
 Cloud 1 overlaps in position with part of the shell, at around $l \sim 131^\circ .1, b \sim 1^\circ .5$, and also in the line-of-sight velocity \citep{Morras98}.
The shell is elongated along the Galactic plane and its size is about $7^\circ \times 3.5^\circ$, approximately constant in the size and position within the $v_{\rm LSR}$ velocity range of --109 to --98 km s$^{-1}$,
 suggesting that the cavity surrounded by the shell has a cylindrical shape \citep{Morras98}.
 
The high-resolution \ion{H}{1} map around Cloud 1 from the CGPS data (see Figure \ref{fig1}) shows that Cloud 1 is associated with an elongated \ion{H}{1} distribution in the same velocity range.
 We estimated the \ion{H}{1} column density, radius, mass, and velocity width of the \ion{H}{1} cloud as $3.6 \times 10^{20}$ cm$^{-2}$, 67 pc, $4.1\times 10^4$ $M_\odot$, and 9.1 km s$^{-1}$, respectively.
\begin{figure}
\epsscale{1.0}
\plotone{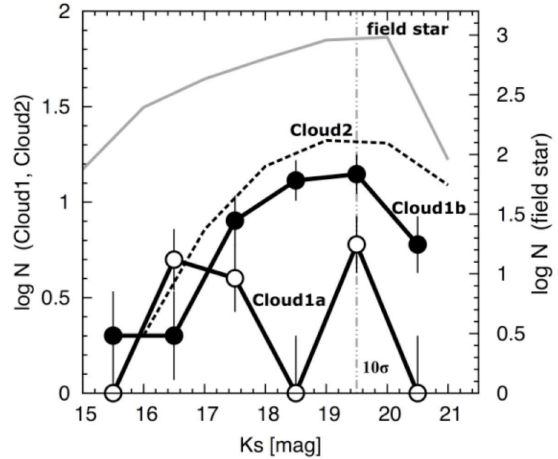}
\caption{$K$-band luminosity function (KLF) of Cloud 1a (open circle), Cloud 1b (filled circle), and Cloud 2 (dashed line) clusters.
The KLF for Cloud 2 is from the left panel of Figure 10 in \citet{Yasui08}.
The error bars show the uncertainties assuming Poisson statistics.
 The gray line shows all of the field stars in the MOIRCS images of Cloud 1.
 The vertical dotted line shows the estimated completeness limit of Cloud 1 data
 (note that the completeness limit for Cloud 2 is about 20 mag).}
 \label{fig4}
\end{figure}
\begin{figure*}
\epsscale{1.0}
\plotone{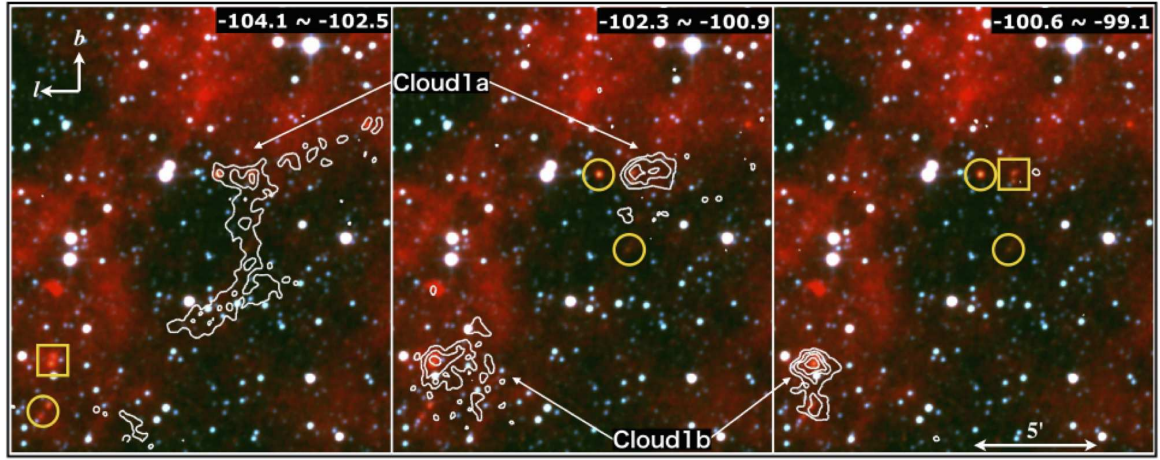}
\caption{$^{12}$CO velocity channel maps for three consecutive line-of-sight velocity ranges in km s $^{-1}$ (from NRO 45 m telescope data) and mid-infrared pseudo color image around Cloud 1.
 The color images are produced by combining the 3.4, 4.6, and 12 $\mu$m images from the $WISE$ data.
 The yellow circles show compact reddened stellar objects in Cloud 1, and the yellow boxes show the Cloud 1a and 1b clusters, which are detected by the 8.2 m Subaru telescope.
 The contour interval is 0.46 K km s$^{-1}$ and range is from 0.46 to 1.82 K km s$^{-1}$.}
 \label{fig5}
\end{figure*}
\begin{figure}
\epsscale{1.0}
\plotone{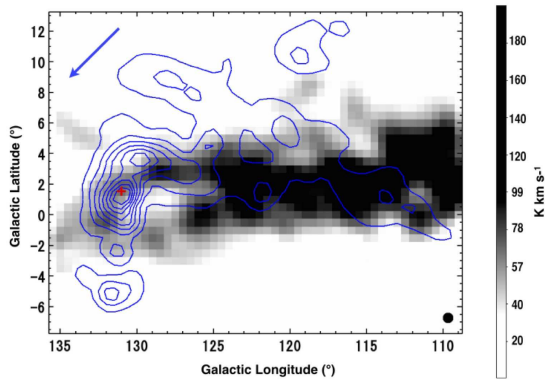}
\caption{\ion{H}{1} clouds in the extreme outer Galactic disk and HVC Complex H seen in a wide field from LAB data.
 While the blue contours show Complex H ($v_{\rm LSR}$ = --229.8 $\sim$ --150.5 km s$^{-1}$),
 the grayscale \ion{H}{1} map shows the Galactic disk at $v_{\rm LSR}$ = --104.1 $\sim$ --98.9 km s$^{-1}$.
 The red cross marks the position of Cloud 1 in the Galactic disk ($v_{\rm LSR}$ $\sim$ --101 km s$^{-1}$).
 The contour interval is 10.8 K km s$^{-1}$ (20$\sigma$) and the range is from 8.5  to 73.1 K km s$^{-1}$ (15$\sigma$ to 135$\sigma$).
 The blue arrow shows the direction of Complex H's motion \citep{Lockman03}.
 The black filled circle shows the beam size of the LAB data ($\sim$ 35$^{\prime}$).} 
\label{fig6}
\end{figure}
\begin{table*}
\begin{center}
\caption{Properties of Cloud 1 and Cloud 2}
\begin{tabular}{cccccccccccccc}
\hline \hline
Cloud & Cloud Mass & Number of Stars& Disk Fraction&Age& $R_{\rm G}$ (kinematic)& $R_{\rm G}$ (photometric)\\
&($10^3 M_\odot$)&&(\%)&(Myr)&(kpc)&(kpc)\\
\tableline\tableline
Cloud 1a&3.0&18&$14\pm 10$ (2/14)&$<$1&22&$\ge$ 19&\\
Cloud 1b&3.5&45&$24\pm 8$ (9/37)&$<$1&22&$\ge$ 19\\  
Bridge&4.5&$\cdots$&$\cdots$&$\cdots$&22&$\cdots$\\  \hline
Cloud 2-S\tablenotemark{a}&8.5&66&$27\pm 7$ (16/59)&0.5--1.0&23.6&19\\
Cloud 2-N\tablenotemark{a}&14&72&$9\pm 4$ (5/52)&0.5--1.0&23.6&19\\
\hline \hline
\end{tabular}
\label{tbl:table2}
\tablecomments{$^{a}$ The mass, kinematic distance, photometric distance of Cloud 2 clouds are from \citet{Digel94}, \citet{Stil01}, \citet{Kobayashi08},
respectively. Other parameters of  Cloud 2 are estimated by  \citet{Yasui06, Yasui08, Yasui10}.}
\end{center}
\end{table*}

\section{Properties of  Cloud 1 clusters}
Because the slope and peak magnitude of KLF vary with age and distance \citep{Muench00},
we estimate those parameters of the Cloud 1 clusters by comparing the observed KLF with that of the young (0.5--1 Myr) 
embedded cluster in the EOG, the Digel Cloud 2 clusters \citep{Yasui06,Yasui08}.
The photometric distance of Cloud 2 has been estimated to be $R_{\rm G}$ = 15--19 kpc ($D$ = 8--12 kpc)
by high-resolution optical spectroscopy of a B-type star MR1 \citep{Smartt96},
which is apparently associated with Cloud 2 \citep{De93};
the shortest and longest distances are based on LTE and nonLTE model
stellar atmospheres, respectively.
Each producing errors are less than 15\% ($D$ = $8\pm$1--12$\pm$2,  $R_{\rm G}$ = $15 \pm$1--19$\pm$2).
In this paper, we adopt $R_{\rm G}$ = 19 kpc ($D$ = 12 kpc) because the nonLTE
model is more likely to
be accurate for stars in the effective temperature regime of MR-1 \citep{Smartt96,Kobayashi08}
and also because $R_{\rm G}$ = 15 kpc has too much large discrepancy with $R_{\rm G}$ = 22 kpc from the kinematic distance.
In Table 2, the properties of the Cloud 1 and 2 clusters are listed.

\subsection{Age}
We estimated the age using the slope of KLF, which is modeled to vary with age \citep{Muench00} and the slope becomes steeper with older age.
\citet{Yasui06,Yasui08} discussed the age of the Cloud 2 clusters by comparing observed KLF and model KLFs of various ages.
They  estimated the ages of the Cloud 2 clusters to be 0.5--1 Myr, and 2 Myr at most, because the model KLF of 0.1 Myr and 2 Myr have gentler and steeper slopes, respectively, 
than the observed KLF of the Cloud 2 clusters (see Figures 7 and 10 in Yasui et al. 2006 and 2008, respectively).
Figure \ref{fig4} shows that the KLF of the Cloud 1b and Cloud 2 clusters have a  similar slope between $K_{\rm S}$ = 16 to 19 mag, and therefore,
we estimated the age of the Cloud 1 clusters to be similar to the Cloud 2 clusters (0.5--1 Myr).
 
 As an additional check of the age, we derived the disk fraction (DF) of the clusters, which is the percentage of cluster members with a optically thick circumstellar dust disk.
 Because the DF is known to decrease with increasing age up to 10 Myr \citep[e.g.,][]{Lada99,Haisch01,Hernandez07,Yasui10}, it is sometimes used for the estimate of the cluster age.
 Following the description in \citet{Yasui09,Yasui10}, we derived the DF of the Cloud 1 clusters using the $H$--$K_{\rm S}$ versus $J$--$H$ color-color (CC) diagram.
 The resulting DF of 24 $\pm$ 8\% (9/37) for detected sources with S/N $>10 \sigma$ suggests that the age of the  Cloud 1 clusters is less than 1 Myr assuming the rapidly decaying DF curve
 in the low-metallicity environment \citep[Figure 1 in][]{Yasui10}.
 Because the ages estimated from both KLF and DF are consistent, we conclude that the age of the Cloud 1 clusters is less than 1 Myr, suggesting that these clusters are truly young embedded clusters. \\

\subsection{Photometric Distance}
The peak magnitude of KLF is sensitive to age and distance of the cluster. For older age and larger distance, the peak magnitude becomes fainter.
 Figure \ref{fig4} shows that the peak magnitude of the Cloud 1b cluster is similar to or fainter than, that of the Cloud 2 cluster.
 In view of the similarity of the age of the Cloud 2 cluster (see Section 4.1), the Cloud 1 clusters are expected to have the same or larger distance than the Cloud 2 cluster.
 Therefore, the distance to the Cloud 1 clusters with $R_{\rm G}$ is suggested to be more than 19 kpc ($D\ge$ 12 kpc),
 which is consistent with the kinematic distance \citep[$R_{\rm G}$ = 22 kpc;][]{Digel94}.
In addition, we tried to estimate the distance to the Cloud1 clusters, assuming that the most luminous star in the cluster is Herbig Ae/Be star of 
3--5 $M_{\odot}$, which is suggested for such small clusters \citep{Testi99, Weidner06}.
 The resultant distance is 8 kpc $\le D \le 21$ kpc (15 kpc $\le R_{\rm G} \le 27$ kpc), which is consistent with the estimated distance by KLF and kinematic distance.
 Therefore, we suggest that the Cloud 1 clusters are located in the EOG region.
  
\begin{figure*}
\epsscale{1.0}
\plotone{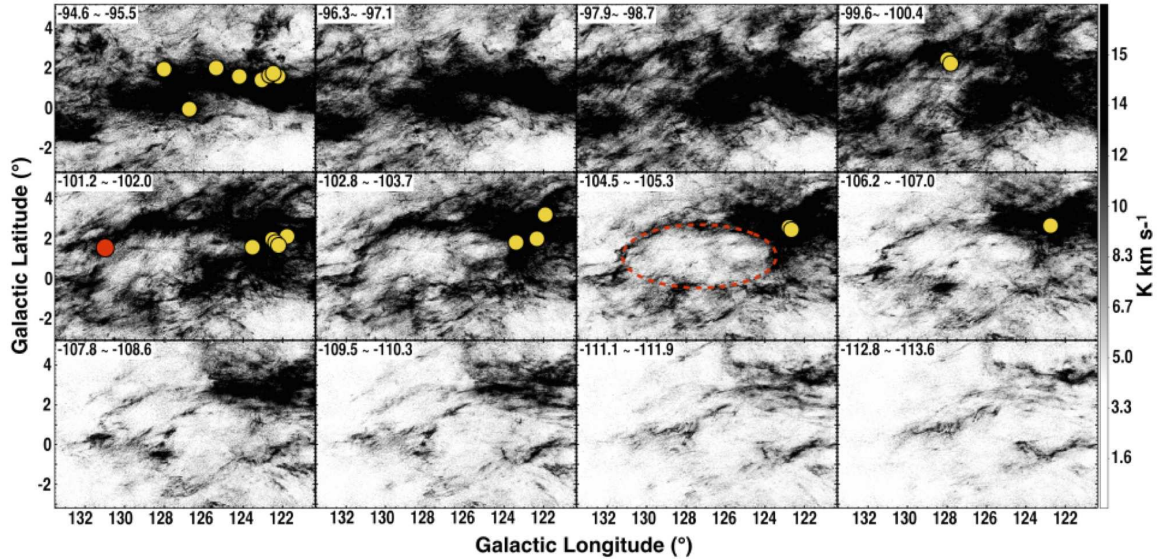}
\caption{\ion{H}{1} channel map from the CGPS data. Every other channel between --94 and --113 km s$^{-1}$ is shown.
 The red filled circle shows the position of Cloud 1, and the dotted red line traces the large shell structure (see Section 3.4, Section 5.1).
 The yellow filled circles show the positions of the molecular clouds, which are identified in the FCRAO data (see Section 5.1 for detail).} 
\label{fig7}
\end{figure*}
\begin{figure*}
\epsscale{1.0}
\plotone{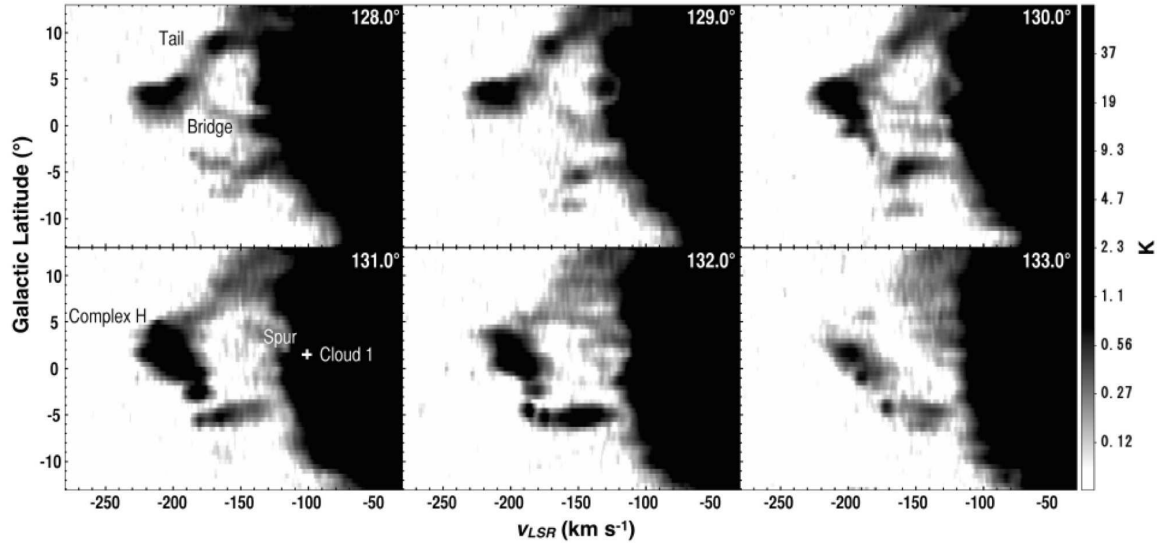}
\caption{\ion{H}{1} velocity-latitude cut through Cloud 1 from $l$ = 128$^\circ$.0 to  133$^\circ$.0 made from the LAB data. The white cross shows the position of Cloud 1.
Besides the probable tidal-interaction ``tail" \citep{Lockman03}, some other ``bridge" or ``spur" features are seen in between the Complex H ($v_{\rm LSR}\sim$ --200 km s$^{-1}$)
and the Galactic disk ($v_{\rm LSR}\le$ --100 km s$^{-1}$).} 
\label{fig8}
\end{figure*}
\section{Possible triggered cloud/star formation in Cloud 1}
Cloud/star formation in a low-density environment, such as the EOG, is of strong interest in connection to the star formation in the early phase of the formation of the Galaxy \citep{Kobayashi08}.
 Because of the low-density, triggered formation as opposed to spontaneous formation may play a crucial role in such an environment \citep{Elmegreen11, Elmegreen12}.
 Triggered star formation on the scale of a molecular cloud is nominally described as follows \citep{Elmegreen98, Elmegreen11},
 (1) stellar pressure (including expansion of \ion{H}{2} region and supernova remnant (SNR) shell), and (2) collision and collapse between two clouds (cloud-cloud collision).
 In the following we introduce the unique environment of Cloud 1 to discuss the possibility that a $large$-$scale$ cloud-cloud collision,
 which was originally proposed by \citet{Morras98}, is the trigger of cloud/star formation of Cloud 1. 
 
\subsection{Large \ion{H}{1} Shell}
Cloud 1 is located on a large \ion{H}{1} shell with a size of 0.8 kpc $\times$ 0.9 kpc at the kinematic distance of $R_{\rm G}$= 22 kpc (Figure \ref{fig6}, see also Section 3.4).
 Therefore, the star-formation mechanism is primarily suggested to be connected to the shell formation.
 In fact, such a large-scale triggered star formation by a super bubble or a super shell is reported for several regions in the Local arm \citep[e.g.,][]{Lee09}, the Perseus arm \citep[e.g.,][]{Sakai14,Lee08}, and the EOG \citep{Kobayashi08}.
In addition to Cloud 1, we noticed several more molecular clouds, which are associated with the shell, from the list of molecular clouds identified in the FCRAO data \citep{Brunt03}.
 Because no molecular cloud is found inside or outside the shell (see Figure \ref{fig7}), we suggest that all those clouds are related to the shell formation.
 We will discuss the star formation activity in those clouds elsewhere in the future.

\subsection{Stellar Feedback?}
First, we discuss the possibility that the shell formation was triggered by stellar feedback.
 \citet{Morras98} estimated the energy required to produce the large \ion{H}{1} shell by a sudden explosion is of the order of $\sim 10^{53}$ erg,
 which requires the combined action of stellar winds and supernova explosions.
 However, a large OB association, which can make such a large shell, has not been detected near Cloud1 \citep{Morras98}.
 To confirm this idea, we reexamined the presence of any prominent sources of stellar pressure inside the \ion{H}{1} shell, using the latest archival data compiled after Morras's work :
 H$\alpha$ images (IPHAS), MIR images ($WISE$), and \ion{H}{1} data (LAB and CGPS).
 However, we could not identify any source and/or structure that traces an OB association or an SNR.
 
 \subsection{HVC Impacting on the Galactic Disk?}
Next, we pay attention to the interaction between Complex H and the Galactic disk, which was first noted by \citet{Morras98} and later discussed by others \citep[e.g.,][]{Blitz99,Lockman03,Simon06},
 to discuss if it can cause the \ion{H}{1} shell formation as well as the molecular cloud/star formation. 
 Based on  \ion{H}{1} data with the Effelsberg 100 m Radio Telescope, \citet{Morras98} suggested the impact of an HVC on the outer Galactic disk, which resulted in the presently
 observed Complex H and the large \ion{H}{1} shell.
 Based on highly sensitive data with the Green Bank Telescope, \citet{Lockman03} paid attention to the ``tail" structure between Complex H and the Galactic disk in a position-velocity (PV) map,
 and suggested that the Complex H is more like a satellite of the Galaxy in an inclined retrograde orbit, whose outermost layers are currently being stripped away in its encounter with the Galaxy.
 However, in a PV map made from the LAB data (Figure \ref{fig8}), we have noticed that some intermediate velocity structures of the ``spur'' and ``bridge", which connect the Complex H to the Galactic disk,
 in addition to the ``tail" structure, and that Cloud 1 appears to be located at the edge of such structures. 
 Although the large-scale tail structure is likely to be formed by tidal force \citep{Lockman03}, the existence of the bridge and spur structures in the PV diagrams as well as the \ion{H}{1} shell structure (Section 5.1) support the
 Morras et al. impacting idea because such structures are predicted by the simulation of cloud-cloud collision \citep[e.g., Figure 2 in][]{Comeron92}.
 
 \citet{Blitz99} posed a major objection to the Morras et al. impacting idea because there is no trace of an impact, such as H$\alpha$ and X ray emission from a strong shock,
 suggesting that Complex H is an extragalactic \ion{H}{1} cloud.
 \citet{Simon06} followed the Blitz's extragalactic idea to present the argument that Complex H is either a dark galaxy in the Local Group, or an example of a cold accretion flow onto the Galaxy.
 However, a similar case of an \ion{H}{1} cloud impacting on the Galactic disk, which does not show any detectable H$\alpha$ or soft X-ray emission, is reported for HVC 306-2+230 by \citet{McClure08},
 who argue that such emissions from associated ionized gas are absorbed by foreground dust and gas in the Galactic plane. 
 The Complex H is also located at low-galactic latitude, and moderately high extinction is measured for the large distance.
 Therefore, the lack of H$\alpha$ and X ray emission does not appear to be strong evidence against Morras' impacting idea.
  
We also considered the timescale of the collision (impact) in relation to the cloud/star formation in Cloud 1.
 Assuming that the \ion{H}{1} peak of the Complex H has a spherical shape with a radius of $R\sim$1.4 kpc ($5^\circ$ at $R_{\rm G}$= 22 kpc) and the relative velocity of $\Delta v \sim$ 100 km s$^{-1}$,
 the estimated dynamical timescale of the collision is $R/\Delta v \sim$ 10 Myr.
 The typical timescale for formation of a molecular cloud is considered to be $\sim$ 10 Myr \citep{Ballesteros07,Gratier12}.
 The lifetime of a molecular cloud as well as the timescale of star formation is also considered as $\sim$ 10 Myr \citep{Mouschovias06}.
 All of these timescales are not longer than the estimated collision timescale.
 In fact, the ages of Cloud 1 clusters are estimated to be $<$1 Myr (see Section 4.1).
 In all possible cases in which the impact triggered (1) both molecular cloud and star formation, (2) only molecular cloud formation, or (3) only star formation in Cloud 1,
 the collision timescale does not conflict with the timescales of subsequent processes.
 
 Therefore, we suggest a possibility that the formation of the Cloud 1 clusters and Cloud 1 itself was triggered by the impact of Complex H on the Galactic disk at $R_{\rm G} \sim$ 22 kpc.
 Further study of this cloud will be very important for revealing the dynamical processes of such triggered formation.
 
\section{Summary}
We present a detailed study of Digel Cloud 1, which is one of the farthest molecular clouds in the Galaxy, with the kinematically determined Galactocentric radius of $\sim$ 22 kpc.
 Our main results are as follows.
 
\begin{enumerate}
\setlength{\itemsep}{-3pt}
\item With NIR imaging and $^{12}$CO mapping that covers the entire Cloud 1, we detected two young embedded clusters located in two dense cores.
\item Using properties of the KLF and DF, we have estimated the age of the clusters to be $<$1 Myr.
\item Using properties of the KLF and the above age, we have estimated the photometric distance of the clusters to be $D\ge$ 12 kpc ($R_{\rm G}\ge$ 19 kpc), which is consistent with the kinematic distance.
\item Based on previous research on Complex H and the latest \ion{H}{1} survey data (CGPS and LAB), we suggest that the impact of HVC Complex H onto the outer part of the Galactic disk
 could be the major trigger of Cloud 1 formation as well as star formation in Cloud 1.\\
\end{enumerate}

We are grateful to the staff of the Subaru 8.2 m telescope, in particular, Ichi Tanaka for his dedicated support during and after the observation.
We are also grateful to the staff of the NRO 45 m telescope.
We thank the anonymous reviewer for a careful reading and thoughtful suggestions that significantly improved this paper.


\end{document}